\documentclass{article}
\usepackage{gastex}
\usepackage{geometry}
\usepackage{url}
\begin{document}

\title{{\bf Erratum : {\sc MCColor} is not optimal on Meyniel graphs}}

\author{{\bf Benjamin L\'ev\^eque,
%\thanks{E.N.S. Lyon.  E-mail:benjamin.leveque@imag.fr}, 
Fr\'ed\'eric Maffray}}
%\thanks{C.N.R.S.  E-mail: frederic.maffray@imag.fr}}}
%\\ \\
%Laboratoire Leibniz-IMAG, 46 avenue F\'elix Viallet,\\
%38031 Grenoble Cedex, France}
\date{}

\thispagestyle{empty}

\maketitle

\thispagestyle{empty}
A Meyniel graph is a graph in which every odd cycle of length at least five has two chords.
In \cite{col} we claimed that our algorithm {\sc MCColor} produces an optimal coloring for every Meyniel graph. 
But later we found
a mistake in the proof and a couterexample to the optimality, which we present here.
{\sc MCColor} can still be used to find a stable set  that intersects all maximal cliques of a Meyniel graph in linear time.
Consequently it can be used to find an optimal coloring in time ${\cal O}(nm)$, and the same holds for Algorithm {\sc MCS+Color}.
This is explained in \cite{stable} but this is equivalent to Hertz's algorithm \cite{her}.
The current best algorithm for coloring Meyniel graphs is the ${\cal O}(n^2)$ algorithm {\sc LexColor} due to Roussel and Rusu \cite{rourus}.
The question of finding a linear-time algorithm to color Meyniel graphs is still open.

%A \emph{coloring} of the vertices of a graph is a mapping that assigns
%one color to each vertex in such a way that any two adjacent vertices
%receive distinct colors.  A coloring is \emph{optimal} if it uses as
%few colors as possible.  The \emph{chromatic number} $\chi(G)$ of a
%graph $G$ is the number of colors used by an optimal coloring.
%

In Algorithm {\sc MCColor}, 
colors are viewed as integers $1, 2,$ \ldots\ At each
step, the algorithm 
selects an uncolored vertex for which 
the number of colors that appear in its neighbourhood is maximum, 
assigns to this vertex the smallest color not
present in its neighbourhood, and iterates this procedure until every
vertex is colored. This algorithm can be implemented in linear time ${\cal O}(n+m)$.

\begin{figure}[h]
\begin{center}\compatiblegastexun
\unitlength=1mm
\begin{picture}(61,30)(0,0)
%\thinlines
%\put(-5,-16){\framebox(90,30){}}
\put(0,0){}

\setvertexdiam{6}
\letvertex A=(5,20)     \drawcircledvertex(A){$a$}
\letvertex col=(-1,20)     \drawvertex(col){$1$}
\letvertex B=(5,10)    \drawcircledvertex(B){$b$}
\letvertex col=(-1,10)     \drawvertex(col){$2$}
\letvertex C=(15,15)   \drawcircledvertex(C){$c$}
\letvertex col=(21,15)     \drawvertex(col){$3$}
\letvertex D=(15,30)    \drawcircledvertex(D){$d$}
\letvertex col=(9,30)     \drawvertex(col){$1$}
\letvertex F=(15,0)    \drawcircledvertex(F){$f$}
\letvertex col=(9,0)     \drawvertex(col){$1$}
\letvertex G=(40,0)    \drawcircledvertex(G){$g$}
\letvertex col=(46,0)     \drawvertex(col){$2$}
\letvertex H=(40,10)    \drawcircledvertex(H){$h$}
\letvertex col=(46,10)     \drawvertex(col){$3$}
\letvertex J=(40,20)    \drawcircledvertex(J){$j$}
\letvertex col=(34,20)     \drawvertex(col){$4$}
\letvertex I=(50,20)    \drawcircledvertex(I){$i$}
\letvertex col=(56,20)     \drawvertex(col){$1$}
\letvertex E=(40,30)    \drawcircledvertex(E){$e$}
\letvertex col=(46,30)     \drawvertex(col){$2$}

\drawundirectededge(A,B){}
\drawundirectededge(A,C){}
\drawundirectededge(C,B){}
\drawundirectededge(C,F){}
\drawundirectededge(C,D){}
\drawundirectededge(C,G){}
\drawundirectededge(F,H){}
\drawundirectededge(D,E){}
\drawundirectededge(E,J){}
\drawundirectededge(J,H){}
\drawundirectededge(H,G){}
\drawundirectededge(G,F){}
\drawundirectededge(I,E){}
\drawundirectededge(I,J){}
\drawundirectededge(I,H){}
\end{picture}
\end{center}
\caption{Counterexample to the optimality of {\sc MCColor} on Meyniel graphs}
\label{contrex}
\end{figure}
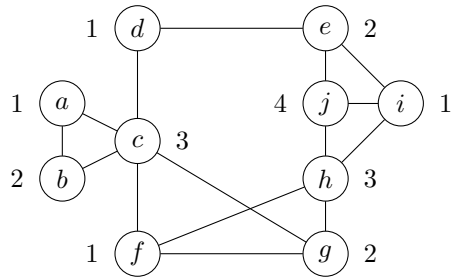

Figure \ref{contrex} shows a counterexample to the optimality of 
Algorithm {\sc MCColor} on Meyniel graphs. The graph is Meyniel
and Algorithm {\sc MCColor} can color
the vertices in the following order, with the given color: $a$-$1$,
$b$-$2$, $c$-$3$, $d$-$1$, $e$-$2$, $f$-$1$, $g$-$2$, $h$-$3$,
$i$-$1$, $j$-$4$. It uses $4$ colors although the graph has
chromatic number $3$.

\

%\newpage

\end{document}